\documentclass[acmsmall,screen,authorversion,nonacm]{acmart}
\usepackage[utf8x]{inputenc} 
\usepackage{tabularx}
\usepackage{booktabs}
\usepackage{listings}
\usepackage[capitalise]{cleveref}
\usepackage{array}
\usepackage{ragged2e}
\usepackage{xcolor}
\usepackage{multirow}
\usepackage{adjustbox}
\usepackage[position=bottom]{subfig}
\newcolumntype{P}[1]{>{\RaggedRight\hspace{0pt}}p{#1}}
\providecommand{\tightlist}{%
  \setlength{\itemsep}{0pt}\setlength{\parskip}{0pt}}
  \definecolor{Gray}{gray}{0.94}

\usepackage{enumitem}

\newcommand{\interviewprefix}{P}
\makeatletter
\newcommand{\interview}[1]{%
  \def\nextitem{\def\nextitem{, }}%
  [\textbf{\@for\el:=#1\do{\nextitem\interviewprefix\el}}]%
}%
\newcommand{\bigquote}[2]{\begin{quote}“\textit{#1}” \interview{#2}\end{quote}}%
\newcommand{\smallquote}[2]{“\textit{#1}” \interview{#2}}%
\makeatother%

\begin{document}

\title{Weaving Privacy and Power: On the Privacy Practices of Labor Organizers in the U.S. Technology Industry}

\author{Sayash Kapoor}
\email{sayashk@princeton.edu}
\authornote{To appear at CSCW 2022. All authors contributed equally. E.A.W. supervised the project. We thank Arvind Narayanan and Andrés Monroy-Hernández for their detailed feedback on a draft of this paper, as well as their feedback and support throughout the project. We also thank the three anonymous reviewers for CSCW 2022 who reviewed our paper and gave us actionable feedback, as well as the 29 labor organizers who generously shared their time and experiences during interviews.}
\author{Matthew Sun}
\email{mdsun@princeton.edu}
\authornotemark[1]
\author{Mona Wang}
\email{monaw@princeton.edu}
\authornotemark[1]
\author{Klaudia Ja\'{z}wi\'{n}ska}
\email{klaudia@princeton.edu}
\authornotemark[1]
\author{Elizabeth Anne Watkins}
\email{ew4582@princeton.edu}
\authornotemark[1]
\affiliation{%
  \institution{Center for Information Technology Policy, Princeton University}
  \state{New Jersey}
  \country{USA}
  \postcode{08540}
}

\renewcommand{\shortauthors}{Kapoor, Sun, Wang, Ja\'{z}wi\'{n}ska, Watkins}
\renewcommand{\shorttitle}{On the Privacy Practices of Labor Organizers in the U.S. Technology Industry}

\begin{abstract}
    We investigate the privacy practices of labor organizers in the computing technology industry and explore the changes in these practices as a response to remote work.
    Our study is situated at the intersection of two pivotal shifts in workplace dynamics: 
    (a) the increase in online workplace communications due to remote work, and 
    (b) the resurgence of the labor movement and an increase in collective action in workplaces---especially in the tech industry, where this phenomenon has been dubbed the \textit{tech worker movement}.
    The shift of work-related communications to online digital platforms in response to an increase in remote work is creating new opportunities for and risks to the privacy of workers. 
    These risks are especially significant for organizers of collective action, with several well-publicized instances of retaliation against labor organizers by companies.
     
    Through a series of qualitative interviews with 29 tech workers involved in collective action, we investigate how labor organizers assess and mitigate risks to privacy while engaging in these actions. 
    Among the most common risks that organizers experienced are retaliation from their employer, lateral worker conflict, emotional burnout, and the possibility of information about the collective effort leaking to management. Depending on the nature and source of the risk, organizers use a blend of digital security practices and community-based mechanisms. We find that digital security practices are more relevant when the threat comes from management, while community management and moderation are central to protecting organizers from lateral worker conflict.

    Since labor organizing is a collective rather than individual project, individual privacy and collective privacy are intertwined, sometimes in conflict and often mutually constitutive. Notions of privacy that solely center individuals are often incompatible with the needs of organizers, who noted that safety in numbers could only be achieved when workers presented a united front to management. 
    Based on our interviews, we identify key topics for future research, such as the growing prevalence of surveillance software and the needs of international and gig worker organizers.
    We conclude with design recommendations that can help create safer, more secure and more private tools to better address the risks that organizers face. 
\end{abstract}



\maketitle


\section{Introduction}

Collective action in the workplace is gaining visibility in the technology industry. Once viewed as both drivers of societal progress and idyllic places to work \cite{barbrook_californian_1996}, tech companies have become more heavily scrutinized in recent years. On the one hand, their role in exacerbating social problems has been the subject of numerous media exposés \cite{buolamwini_gender_2018, isaac_how_2017, florida_tech_2017}. On the other, numerous cases of discrimination and harassment in tech companies---including racism and anti-Black discrimination \cite{bhuiyan_black_2020}, sexism and sexual harassment \cite{fowler_reflecting_2017, wakabayashi_how_2018}, and anti-Trans discrimination \cite{denisco-rayome_transgender_2019}---have brought workplace issues within these companies to the forefront. In addition, the definition of who is considered a tech worker has been expanding, leading to tech workers across the class spectrum finding common ground in their struggles for a better workplace. This has led to an growing number of attempts at collective action in technology companies---ranging from petitions, demonstrations, and walk-outs to unionization \cite{tarnoff_making_nodate}.

Workers face numerous challenges when attempting to take collective action at their workplaces. Employers often retaliate against labor organizers even when it is illegal to do so \cite{lafer_fear_2020, silver-greenberg_fired_2020}. Rank-and-file labor organizers assume significant risks of retaliation via demotions, firings, and other forms of disciplinary action as a result of their organizing efforts.
The increase in workplace surveillance, especially of online communications, heightens these risks as organizers' location, messages, and online activity can be continuously monitored by their employer \cite{cyphers_inside_2020, masoodi_workplace_2021}. 
The COVID-19 pandemic has also drastically altered the nature of communication between workers. Especially in the computing technology industry, more people have transitioned to remote work, resulting in more of their communications taking place online \cite{yang_effects_2021, brynjolfsson_covid-19_2020}. 
This increase in online communication and surveillance has led to employers gaining much more access to the communications of their workers \cite{aloisi_essential_nodate}. Since organizers must often organize co-workers with varying levels of familiarity with technical tools, when using online communication and collaboration platforms they need to carefully balance privacy on the one hand, and usability and reach on the other.

Recent work in usable information security has focused on how groups with a heightened need for privacy navigate online communication and collaboration \cite{mcdonald_its_2021,havron_clinical_2019,owens_x201cyou_2021,simko_computer_2018,mcgregor_investigating_2015}. These efforts highlighted the shortcomings of the traditional information security models and spurred the development of new software tools and practices for these populations \cite{lerner_confidante_2017}. However, no study has as yet explored the privacy models of labor organizers in the computing technology industry and the change in these practices as a response to remote work. 

We study the privacy practices of labor organizers in the computing technology industry, including workers organizing a union or other collective actions such as petitions in their workplace. We address this specific population for three reasons. First, as software continues to ``eat the world'' \cite{andreessen_why_2011}, more and more jobs are becoming mediated by technology \cite{yang_effects_2021}. Documenting the privacy practices of tech workers may have benefits for understanding collective action in other industries whose dominant practices, cultures, and organizational mindsets may come to resemble those of today's computing technology industry. Second, as researchers and organizers who have worked in this sector, our research team felt well-positioned to perform this specific evaluation. Finally, by taking a relatively expansive definition of who constitutes a tech worker (\Cref{sec:inclusion}), we identify commonalities across workers and also point to subgroups whose needs deserve special attention in future work (\Cref{sec:future_work}).

Our primary research questions are: 
\begin{itemize}
    \item \textbf{RQ1}: How do labor organizers in the computing technology industry assess and respond to risks to privacy?
    \begin{itemize}
        \item \textbf{RQ1-a}: What are the specific goals and needs of organizers in the computing technology industry?
        \item \textbf{RQ1-b}: What are the risks that organizers face to these goals and needs, both as a collective and as individuals?
        \item \textbf{RQ1-c}: How do organizers respond to these risks, and what trade-offs do they weigh when making these decisions?
    \end{itemize}
    \item \textbf{RQ2}:  What effects has the shift to remote work and online communication had on this community's practices?
\end{itemize}

The goal of \textbf{RQ1} is to understand how labor organizers evaluate trade-offs and make decisions in response to risks. The process of taxonomizing the goals and risks of a particular situation or context and creating a plan to address them is called \textit{threat modeling}, or \textit{risk assessment}, in the field of information security~\cite{electronic_frontier_foundation_your_2014}.
The various sub-parts of \textbf{RQ1} help us understand how organizers assess and respond to risks when organizing. 
The goal of \textbf{RQ2} is to understand how labor organizing and collective action has changed in the shift to digitally mediated workplaces, especially in response to the COVID-19 pandemic.

We conducted 29 semi-structured interviews with U.S.-based labor organizers in the computing technology industry in order to gather data that would help us to answer these research questions. We wanted to gain insights into participants' privacy practices to understand their threat models, the impact of various digital tools on their perceptions of privacy, the impact of remote work and online communication and collaboration on their organizing, and the impact of social relationships and community guidelines in creating a safe organizing unit. Through asking organizers about their direct experiences in this space, we are able to examine \textit{actualized} risks, in addition to presumptive or assumed risks. 

Following our findings (previewed in the next section), we conclude by offering a set of concrete recommendations for designers of digital communication and collaboration technologies. These recommendations include providing clear documentation on the extent of administrator access to files and messages, searchable archives, and enhanced functionality for collective decision-making. While no technological intervention can or should remove the human element of organizing, we see significant opportunities for practitioners in these areas to better facilitate the protection of the individual and collective privacy of workers---especially since digital tools are often built to serve employers and management rather than the employees working for them.

\subsection{Overview of our findings}
We summarize our results in Tables \ref{tab:results} (\textbf{RQ1}) and \ref{tab:resultsRQ2} (\textbf{RQ2}). Some of our main findings include:
\begin{itemize}
    \item The most commonly experienced risks include pervasive retaliation on the part of the employer as well as lateral worker conflict. The extent to which digital technology aided or hindered risk mitigation depended upon the threat actor. 
    \item When the source of the threat was from management, organizers made use of non-company-owned tech and strict access controls to minimize the possibility of information leakage. 
    On the other hand, mitigating the risks related to worker conflict required robust practices of community management, building social trust, and emotional care work. These practices were especially important in a remote organizing environment, where building trust and moderating discussions are particularly difficult. Organizers observed that in many cases, organizing effectively while protecting safety was not necessarily compatible with adhering to strict digital security practices (e.g., ``just use Signal''). 
    \item Organizers conceive of risks both at the individual and collective level, describing these as both in tension and mutually constitutive. Individual risks included job loss, but also burnout, loss of privacy, and harassment. Examples of risks to the collective included employer discovery of internal strategy, union-busting behavior, and vilification of the collective as greedy or antagonistic. 
    \item Managing these categories of risk was not zero-sum but did require significant labor on the part of organizers to reconcile the needs of individuals with the needs of the collective. Here, we see opportunities for future research to engage with more expansive definitions of safety to encompass both individuals and the collectives of which they are a part.
\end{itemize}


\section{Related Work}

\subsection{Privacy risks and practices of specific populations} 

It is becoming more common in studies of privacy practices in digital communication to focus on the needs of specific at-risk populations. Their experiences can provide insights into holes in digital security guidance and issues with our privacy infrastructure, and enable design recommendations that draw on the insights of their specific needs, contexts, and characteristics. Several at-risk populations have been the studied for their privacy practices and needs, including sex workers~\cite{mcdonald_its_2021}, victims of intimate partner violence~\cite{havron_clinical_2019}, incarcerated people~\cite{owens_x201cyou_2021}, refugees~\cite{simko_computer_2018}, and journalists~\cite{mcgregor_investigating_2015}.

Taken together, this body of work suggests that the usability of common privacy tools is not only impacted by their design, as noted by \citet{whitten_why_1999}, but also by the social, economic, and professional contexts of their users. Moreover, an individual's privacy often involves complex dependencies on other individuals in their networks, such as domestic partners, case managers, and clients; these relationships typically involve power differentials within an organizational or social hierarchy.

Inspired by this tradition, we conduct interviews with a specific group---labor organizers in the computing technology industry---to shed light on the risks its members experience, and how digital technologies are involved in strategies to mitigate and address those risks. Individuals involved in collective action at their workplaces face immense personal risks to their jobs, which affects their financial security, access to healthcare, and ability to support their loved ones. In addition, since labor organizing is a collective rather than individual project, notions of privacy that center individuals may be incompatible with the needs of organizers. This is unlike other at-risk populations where the primary resolution of privacy risks can be centered around better tools for individual privacy. Through our current work, we address how labor organizers come up with practices around privacy-enhancing tools and how they negotiate which practices and tools to adopt as a collective~(\textbf{RQ1}). 

\subsection{Privacy risks in digital communications at work} 

Prior research examining the interaction between digital tools and work practices recognizes that communication technologies often favor "managerial prerogatives" which seek to divide and control labor \cite{fox2020worker,ehn1988work}. Management's access to computation tools has also increased their access to disparate domains of work tasks, making work easier than over to "measure, model, and evaluate" \cite{khovanskaya2019tools} and thus subjugate to managerial oversight. The share of employers who report digitally monitoring employee behavior has increased steadily since the 1990s \cite{nord_e-monitoring_2006}. As the era of big data and machine learning has entered the mainstream, companies have sought to leverage ever-growing streams of data to increase corporate profits---including to better monitor employees. For example, \citet{levy_privacy_2018} introduce the concept of ``refractive surveillance,'' in which data originally intended for tracking consumers in the retail sector becomes re-purposed as a tool for controlling, measuring, and tracking employee behavior. The shift from in-person to remote work amid the COVID-19 pandemic threatens to exacerbate these conditions. Privacy risks are conceptualized and acted upon differently in face-to-face versus remote work, and the widespread reliance on remote work amid the pandemic, we fear, will make it more difficult for workers to resist such managerial prerogatives. 

Workers have faced tensions in how to manage the visibility of their work within their organizations. Khovanskaya et al., in their work on data practices among union organizers \cite{khovanskaya_bottom-up_2020}, describe a series of studies examining information visibility in nursing. They describe that visibility in some instances has led to a gains in power and recognition \cite{bowker1996infrastructure}, yet for others led to loss of autonomy and discretion \cite{star1999layers}. Journalists also wield comparative visibility as a method for enhancing privacy, relying on face-to-face communications for risky or sensitive conversations in order to protect themselves and their sources. Strikingly, \citet{mcgregor_would_2017} found that all participants reported using face-to-face conversation as a means of avoiding a written record for privacy. This reliance on face-to-face communication, in part, motivated the current work and our interest in privacy enhancement amid the COVID-19 pandemic: what strategies do at-risk populations employ when digital methods are the only means of communication?  

We aim to better understand the tools and practices that enable labor organizers to feel safe when organizing online despite the risks associated with digital communication, especially as a result of the COVID-19 pandemic. Prior work on professional field organizers has investigated that access control around member data in unions and highlighted the difficulty of structuring data in a locally useful way \cite{khovanskaya_bottom-up_2020}. Our study builds on this work by analyzing the data strategies workers who are already embedded within organizations and are new to organizing efforts instead of focusing on professional field organizers. 

While previous work gives us an insight into how threats to privacy are operationalized in the workplace, we outline strategies that enable workers to continue organizing despite these risks. In addition, collective wisdom about organizing in the past has centered in-person communications, but does not address online labor organizing in much detail~\cite{mcalevey_no_2016}. We aim to elucidate this point by outlining how labor organizers in tech negotiate between risks in digital communications at work, especially with the shift to remote work as a result of the COVID-19 pandemic (\textbf{RQ2}).


\section{Background}

Our study builds on a long history and literature of labor organizing and tactics for building collective power~\cite{mcalevey_no_2016}. Here, we provide a high-level overview of some salient distinctions between individual and collective action, as well as ongoing work to support collective action in the tech industry.

\subsection{Collective vs. individual action in the computing technology industry}

Actions designed to shift power from capital to labor can take many different forms. In this section, we distinguish between individual action versus collective efforts. Our study focuses on collective action, which relies on grassroots organizational infrastructure built by a collective of workers, and resilience through worker solidarity in large numbers. This includes union organizers and organizers of other actions such as petitions, walkouts, strikes, and protests within workplaces.

We distinguish collective action from individual action, and focus on the former. Whistle-blowing, one of the most visible forms of individual action, has become more common in the computing technology industry, as individual employees or former employees with moral disagreements with their workplaces air their concerns publicly. Examples range from Sophie Zhang and Frances Haugen, both former employees at Facebook concerned about the social impact of the company's products, to Ifeoma Ozoma and Susan Fowler, who blew the whistle on discrimination and harassment at Pinterest and Uber, respectively \cite{hao_she_2021, pelley_whistleblower_2021, woo_tech_2021, fowler_reflecting_2017}. After the fact, whistle-blowers often experience targeted surveillance or retaliatory behavior by their former employers. For example, Fowler reported being followed by private investigators hired by Uber executives after publishing her blog post \cite{lopatto_susan_2020}.

Prior work also highlights a growing anxiety among activists that the shift towards individualism in advocacy spaces can hinder collective action and resilience \cite{aouragh_lets_2015}. Whereas whistle-blowers tend to be seen as \textit{lone wolves} who solely bear the risks and rewards of their decisions to come forward, collective action involves threats to privacy that are distributed among many individuals whose diverse perspectives and constraints shape the decisions made by the collective. The focus of our work is on collective decision spaces that require complex modes of reconciliation to address tensions.

\subsection{Toolkits, events and projects for organizing the computing technology industry}

There are a number of initiatives aimed at supporting organized collective action in the tech sector. For example, \textit{\#NoTechForICE} targeted computing technology companies who had contracts with U.S. Immigration and Customs Enforcement through both internal employee protest and student pledges to withhold labor \cite{denham_no_2019, noauthor_students_nodate}. The \textit{Contract Worker Disparity Project} aims to shed light on compensation and benefit inequity between contract workers and full-time employees in the tech industry \cite{noauthor_contract_nodate}. Media projects include \textit{Collective Action in Tech}, a space for documenting collective actions in the tech industry and sharing resources \cite{tarnoff_about_2020}, and \textit{Digital Worker Inquiry}, an event that took place in October 2021 that showcased different worker-led data projects for understanding and resisting their working conditions \cite{noauthor_digital_nodate}. 
In recent years, tech-specific unions have also emerged. In January 2020, the Communication Workers of America launched the \textit{Campaign to Organize Digital Employees (CODE-CWA)}, a network of worker-organizers in the tech, game, and digital industries \cite{noauthor_campaign_2019}.  In January of 2021, Office and Professional Employees International Union (OPEIU) launched \textit{Tech Workers Union Local 1010}, a union run by tech workers to support workers organizing at their companies and to build solidarity across workplaces \cite{noauthor_about_nodate}.

We provide this non-exhaustive list of projects to emphasize the important work that is occurring outside academia on ensuring the privacy of organizers. While the academic literature on the privacy needs of labor organizers may be thin, these projects show how tech workers are already creating knowledge to protect themselves and build power. Our work is built upon this existing foundation of worker-led initiatives.


\section{Methods}

To answer our research questions, we conducted 29 qualitative semi-structured interviews in combination with a quantitative demographic survey. These interviews were held in October-December 2021. Qualitative interviews provide an instrument to gather data on participants' perceptions of their own risks, environments, insights into the cognitive and sociological factors which motivate their behaviors, and their own perceptions of the relationships between these behaviors and subsequent results. We used an optional survey instrument to assess the demographic representation of our participant pool.

\subsection{Inclusion criteria} 
\label{sec:inclusion}
In our study, we interview U.S.-based tech workers who had organized collective action at a non-unionized workplace in the past five years. 
We describe our definitions for each term and the limitations that result from these criteria. 
Our definition of \textit{tech worker} included both workers who had a technology role at non-tech companies (e.g., software engineers or IT workers) as well as anyone who worked at a technology company irrespective of their role (e.g., marketing staff at a company whose main product is a digital technology). This definition includes full-time employees, gig workers and contractors. While we tried to ensure that our sample captured as many possible viewpoints from different positions in the computing technology industry, our sample skewed towards full-time employees rather than gig workers and contractors. 
Our definition of \textit{collective action} was quite broad and included workers who had organized their co-workers to perform collective actions to affect a company policy or decision. For instance, this would include organizing a collective to issue a petition, perform a walk-out, or other forms of work stoppage in addition to attempting to form an NLRB-recognized labor union. We define an organizer as someone who talked to at least one other person with the intent of assessing or securing support for the collective action.
We focus on organizing at non-unionized workplaces. We recognize that unionized workers may also regularly organize their co-workers to participate in actions, in particular during regular contract negotiation. Since there are fewer established unions in the computing technology industry, and we believe the risks that organizers face in a unionized workplace may be different than at non-unionized workplaces, we decided to focus on the latter.
We limit our study to recent organizing activity (in the past five years) because we are interested in studying how recent shifts towards online or digitally mediated workplace communication have affected organizing. In particular, the COVID-19 pandemic led to a growing number of companies, especially in the tech sector, to adopt remote-friendly policies.

\subsection{Interview procedure} 
In this section we discuss in detail our interview procedures, from recruitment through the interview itself and optional survey. These were also reviewed and approved by our university's Institutional Review Board (IRB). 

\subsubsection{Recruitment} 
We recruited participants using four mechanisms: through our personal networks, through a social media post \cite{elizabeth_anne_watkins_phd_are_2021}, through cold messages to public social media accounts, and through snowball sampling using recommendations from the participants we interviewed. We asked individuals to self-identify whether they fit in the inclusion criteria we outlined in \Cref{sec:inclusion}.

\subsubsection{Interview instrument}

Our interviews started with a discussion of the participant's role at the company where they were involved in collective action. We then moved to questions about specific collective actions that the participant was involved in organizing, and the processes through which they conducted these actions (\textbf{RQ1-a}).
We then asked the participant about their perceived risks, both to the collective and to the individual (\textbf{RQ1-b}). 
We moved on to ask how they managed risks collectively and individually, and how they managed various trade-offs when designing policies, processes, and choosing digital tools for communication and collaboration (\textbf{RQ1-c}). 
Finally, we explored how organizing remotely had affected the participant's experience with collective action, and if they had been organizing before the pandemic, what differences they had observed before and after the pandemic (\textbf{RQ2}).
The complete interview instrument is attached in \Cref{sec:interview_instrument}.

\subsubsection{Conducting the interview} Participants were interviewed over a Zoom video call. The interviews lasted between 90-120 minutes, and participants were remunerated with 50 USD prepaid debit cards after the interview. Two members of the research team interviewed each participant---one led the interview while the other transcribed the interview and asked questions of interest at the end of the interview.

\subsection{Analysis}
Our approach to qualitative data analysis focused on structural, conceptual coding via abductive analysis, meaning that we iterated between the high-level concepts we observed in our collected data and related literature in threat modeling \cite{saldana_coding_2021, tavory_abductive_2014}.
After 15 interviews had been completed, all members of the research team read over all of the interview transcripts, and then convened to conduct an initial open coding session and identify and discuss themes in the interviews that had been conducted so far. We identified large conceptual themes, under which we categorized specific instances or manifestations that had been mentioned by interview participants. Through this discussion the team created a set of initial conceptual codes which aligned with research questions. 

When nearly all interviews had been completed, the team met again. Prior to this second group meeting, each researcher was assigned a set of six interviews to read for which they had been neither a primary interviewer nor note-taker. This ensured that each interview was "read" by at least three people, including the initial two researchers who conducted the interview as well as an additional reader. The second discussion focused on axial coding. This session determined the significance of the initial codes and the relationships between the categories they designated,  elevating those concepts most significantly representative of the teams' research interests. During this second session, it was determined that a broadened threat model framework presented a comprehensive and accessible way to taxonomize our conceptual codes. One researcher used these frameworks to draft a codebook, after which they and two other researchers conducted a round of focused coding on a subset of randomly selected interviews to identify and eliminate any redundancies between codes, or identify and eliminate any codes which were overly vague or ill-defined. They presented their findings to the group, after which the codebook was refined again, another researcher joined the coding effort, and another subset of interviews were randomly selected and coded to ensure reliability of the codes and the categorization and reporting of results.  

We specifically did not measure inter-rater reliability. As McDonald et al. argue, calculating inter-rater reliability is a methodologically poor match for interpretive research, when developing codes are a process for determining themes, rather than the product of the research themselves \cite{mcdonald_reliability_2019}. Hence, because our coding process is instrumental to formulating and theorizing a diverse set of themes, we choose not to calculate inter-rater reliability as part of our analysis. Further, we have chosen not to rely on quantitative counts of instances to describe our findings. Because qualitative research is drawn from an interpretive paradigm, and because our coding taxonomy (i.e. threat model frameworks) is procedural and not categorical, we found that data as provided by participants were subject, in a number of instances, to multiple complementary codes (i.e., the same instance could be coded as indicative of multiple points within a threat model, concurrently describing, for example, risks and risk mitigation strategies).

\subsection{Demographic survey} 
For the participants we interviewed, we conducted an optional demographic survey to study attributes such as race, gender and visa status, a design based on prior demographic surveys from Pew Research Center \cite{noauthor_pew_2015}.
We sent the survey forms via email after each interview; all questions were optional to allow for completely voluntary self-disclosure. The survey data was collected via end-to-end-encrypted forms~\cite{noauthor_cryptpad_nodate}. The survey instrument is attached in \Cref{sec:survey_form}.

\subsection{Ethics, privacy, and safety considerations} Given the sensitive nature of our interviews and since participants were taking on the risk of employer retaliation by participating in our interviews, we took several measures to reduce risks for our participants. 
We sought to not only minimize the sensitive data we as researchers held about each interview, but also to minimize the data accessible to any digital intermediary. We did not know of any other research data policy that had these standards for holding digital interview data, so we drafted a detailed, custom policy and had it reviewed by labor organizers as well as our institutional IRB. At a high level, our privacy policies were as follows:

\begin{enumerate}
    \item \textit{No recordings.} We did not record the interview to avoid storing audio data from the interview. Instead, each interview involved two members of our research team, with one of us transcribing the interview. This also meant that the interview was not accessed by any third-party (for example, transcribing services) and remained solely within our research group.
    \item \textit{Anonymizing Personally Identifable Information (PII).} We removed all PII from the interview transcripts. In addition, we removed references to individuals, companies, or organizations that could be tied to the interview subject, and redacted terminology that may have been specific to a particular company or organizing team.
    \item \textit{Password-protected transcript files.} All interview transcripts were stored in password-protected files on a cloud storage platform that was accessible only to the research team. The passwords were only known to the research group, so even the cloud storage platform would not have access to the interview transcripts.
    \item \textit{Data retention policy.} After sending the interview subject their compensation for the interview, we removed any records that linked the names or contact information of the participants with their interview transcript.
    \item \textit{Communication policy.} Any discussions of individual interviews were conducted solely between the research team in-person, over an end-to-end-encrypted video call, or via an end-to-end-encrypted messaging platform with disappearing messages.
    \item \textit{Reporting participant labels anonymously.} To avoid identifying participants based on several quotes attributed to them in the paper, we change the participant labels in each section of the results. For example, a participant might get the label [\textbf{P23}] in \cref{sec:risk}, [\textbf{P14}] in \cref{sec:remote-work}, and [\textbf{P8}] in \cref{sec:other-trends}. This further reduces the chances of identifying participants by combining information available about them from several quotes. 

\end{enumerate}

The only data other than the anonymized transcript we retained were in aggregate, such as results from the optional demographic survey and employee job type (contractor, full-time, or platform-worker).

\subsection{Positionality} \label{sec:positionality} We undertake this project not as outside observers, but as people enmeshed within the web of technology, capital, and labor \cite{salehi_we_2015}. Through the lens of standpoint theory, we recognize that we research and write from a "particular, historically specific, social location"\cite{harding2004feminist}. Our authors include former tech workers who have experienced, heard about, or been witness to labor organizing efforts in their workplaces. As knowledge workers in academic computer science, all of us have stakes in advocating for better working conditions, as well as a responsibility to steer our research in ways that promote the employees’ privacy as they exercise their rights to collective action.
Our positionality also influenced the kinds of information we were able to glean through our interviews. Our own unconscious biases and positionality as researchers at an 
academic institution likely influenced participants' willingness to speak with us as well as what and how they chose to share during their interviews. Though some of these biases can be corrected for through reflexivity and adherence to best practices, it remains important to recognize their influence on the project.

\subsection{Limitations}
\subsubsection{Inclusion criteria}
We recognize that our inclusion criteria limits our scope of view. The recent surge in labor organizing is not at all limited to the computing technology industry, and many organizers not included in the above criteria may find their personal experiences reflected in our findings. All workers face many of the same risks when organizing, and we recognize that organizers at unionized workplaces must also regularly organize their co-workers to participate in collective action.

Our inclusion criteria also has blind spots. For instance, we cannot make any general claims about how the labor organizers in our sample reckon with physical security concerns (for instance, from physical surveillance by supervisors) because every worker we interviewed was organizing remotely, or at some point shifted to organizing remotely.
Additionally, while workers outside of the U.S. were not included in our interviews,  many of the essential forms of labor that fuel the technology industry are primarily located outside of the U.S. \cite{sambasivan_seeing_2021}. In a world where such labor is unevenly distributed by geography, we hope that follow-up work explicitly focuses on workers outside of the U.S., particularly in the Global South.
Furthermore, we caution against generalizing our results even within our inclusion criteria. While the participants in our study represent a diversity of viewpoints, roles, and demographic groups, it would be inaccurate to say that they are representative of all workers in the U.S. computing technology industry at large. 

\subsubsection{Imperfect information due to human error and legal considerations.}

Because we did not record interviews, we relied on transcripts typed by one of the interviewers during the video call. It is possible that certain words or details were transcribed incorrectly. We are also unable to review transcripts for accuracy. In addition, as part of our interview procedure, participants were told to use their best judgment when disclosing potentially confidential information about the companies for which they worked. We have no way of knowing whether some details were withheld from us on the basis of legal concerns.

\subsubsection{Population bias.}

As we mention in \Cref{sec:positionality}, our positionality as academic researchers and the subjective experiences of labor organizers will inevitably have an effect on the composition of our interview subjects. Our survey (\Cref{sec:survey_results}) showed us that our population skewed young and highly educated, which could demonstrate a bias in our network-based sampling method. In addition, we suspect our social media and public sampling method may have introduced additional biases. For instance, the majority of our interview subjects were full-time employees. Under U.S. labor law, contracted workers enjoy very few workplace protections, and therefore may not feel as comfortable being public about organizing, or participating in voluntary academic research. We acknowledge these limitations of our results, which we hope can be addressed in future work.


\section{Results}

From the interviews we conducted, we collect and describe the needs and goals of organizers, the risks they face to both individual and collective privacy, and the strategies they use to mitigate those risks.
Our participant and workplace composition are summarized in \Cref{tab:composition}. In total, we interviewed 29 organizers at 17 different workplaces. To prevent oversampling from one particular company, we did not interview more than 3 organizers at the same workplace. 24 were full-time employees, 2 were contracted workers, and 3 were platform-workers. Note that \textit{platform} workers' employment is primarily mediated through a gig work software platform, whereas contractors had less transient workplaces and generally worked full-time schedules. 19 of the participants were involved in a campaign towards an NLRB-recognized labor union, and 10 of the participants were involved in organizing co-workers in order to perform other collective actions at their workplace. We note that under U.S. labor law, many of the workers we interviewed (i.e. contractors) were not able to campaign for an NLRB-recognized labor union. Of the workplaces where participants organized, three were 501(c)(3) non-profit organizations, and the other 13 were for-profit companies. Eleven of the for-profit companies' core products were digital platforms or applications, so we therefore classified them as \textit{technology} companies.
 
\begin{table}%
  \centering
    \begin{adjustbox}{width=\textwidth,center}
    \subfloat[][Breakdown of participant employment status during their organizing effort.]{
    \begin{tabular}{ll}
    \toprule
    \textbf{Employment status} & \textbf{\#Participants} \\
    \midrule
    Full-time employee& 24 \\
    Platform-worker & 3 \\
    Contractor & 2 \\
    \bottomrule
    \end{tabular}
    }%
    \qquad
    \subfloat[][Breakdown of the companies where our participants organized collective action.]{
    \begin{tabular}{lll}
    \toprule
    \textbf{Company type} & \textbf{\#Companies} & \textbf{\#Participants}\\
    \midrule
    Technology (for-profit) & 11 & 23\\
    Other (for-profit) & 3 & 3\\
    Non-profit & 3 & 3\\
    \bottomrule
    \end{tabular}
    }
    \end{adjustbox}
    \caption{The workplace and job type composition of our interview participants.}%
    \label{tab:composition}%
\end{table}
 
\subsection{Needs and goals of organizers (\textbf{RQ1-a})} \label{sec:needs}

Needs and goals refer to the artifacts that labor organizers want to protect and preserve during their organizing efforts. This is a broader framing of the concept of \textit{information assets} in threat modeling---as demonstrated in related literature, \textit{assets} alone cannot encapsulate the needs of particular groups that must also work to achieve certain organizational or political goals~\cite{aouragh_lets_2015,electronic_frontier_foundation_your_2014}. Here, we describe the salient needs and goals that participants described during their organizing effort.

\subsubsection{Needs for the campaign} Although the type of collective action (i.e. walk-out, petition, or organizing a labor union) varied across our interviews, the methods used were similar. In all of our interviews, organizers reached out to as many workers as possible, or in some cases, every single worker. They held one-on-one meetings with each of these workers to discuss workplace conditions, assess their willingness to help outreach and organizing efforts, and assess their support for the collective action. Especially after workplaces went remote, organizers also regularly checked in with co-workers about their working conditions and experiences.

\subsubsection{Collective information} In all of our interviews, organizers kept some form of documentation around worker concerns, working conditions, and outreach status. This included names of fellow organizers and supportive co-workers, i.e. member lists, records of organizing efforts, as well as documentation of institutional memory, including workers' shared accounts of their motivations to organize, and documents containing organizing strategy.

\subsubsection{Individual relationships and assets} Throughout this process, in addition to organizational memory and collective data, individual organizers or participants in the collective action wanted to protect their own information, relationships, or financial security, to the extent possible. Organizers spoke about using discretion when sharing sensitive stories or information about co-workers that may relate to their working conditions \interview{3,11,22,27}. Individuals also wanted to protect their relationships with co-workers as well as with their managers or employers; it was often important for organizers to not be perceived differently within their workplace because of their involvement in collective action \interview{11,14,17}. Finally, in almost every interview we conducted, organizers were concerned about whether perceptions of their organizing activity could influence their job security.

\subsection{Risks faced by organizers (\textbf{RQ1-b})} 
\label{sec:risk}
Organizers' evaluations of risks include both the actors (e.g., employers, co-workers) from whom organizers face threats, as well as scenarios where any of the needs or goals listed above can be put at risk. These risks were presented as dynamic and ongoing evaluations, with one participant stating, \smallquote{[after some time organizing,] it became clearer through observation what kind of tactics the company is willing to engage in. We had a better understanding of union-busting in general and that we need spaces to have these conversations that are more private or more protected}{10}.
In our interviews, the risks faced by organizers fell under two main headings: risks to the individual organizer, and risks to the campaign. 

\subsubsection{Risks to the campaign or collective effort} These include any events that could make it less likely for the organizing effort to succeed. While most of these risks originated from employers, some also originated from co-workers. We highlight a few examples below:

\begin{itemize}
    \item \textit{Employer finds out about general strategy.} Organizers were wary of their employer finding out, through information or access leaks, the existence of or general strategy for collective action that they were going to deploy. Many considered this a significant risk to the collective action \interview{14,17,21,27}. One participant related, \smallquote{we're doing critical organizing, and we don't want people who may have leaked information}{14}. They felt that this could lead to broad efforts by the employer to influence the campaign, interfere with their efforts, and eventually result in an unsuccessful campaign. There were several cases where the entire organizing effort was pre-emptively leaked to management \interview{2,7,8,11,23}. In one case, organizers saw a rise in co-ordinated anti-collective rhetoric, and the company started to pre-emptively address previously ignored organizer concerns just before the collective went public. Organizers typically waited for as long as possible before making their campaign public to executives.
    \item \textit{Employer hires a union-busting firm.} One of the most common responses when an employer finds out about organized collective action is to hire a union-busting firm, or anti-union legal counsel. These firms attempt to break up unionization or other collective activity inside companies using several tactics to create a culture of fear of the collective and sowing distrust amongst workers \cite{logan_union_2006}. Many participants' companies ended up hiring anti-union legal counsel or union-busting firms \interview{3,9,11,14,17,29}. One participant expressed that hiring a union-busting firm was a major risk to the collective effort if the employer found out about their campaign, since they can use their unfettered access to the company's communication channels and employees to sow divisions within the workplace and severely hurt the effort. Yet another related, \smallquote{union-busting fucks up company culture. A nightmare honestly. I don’t have words to describe it. It's like psychological warfare that your workplace is throwing at you.}{3}
    \item \textit{Lateral worker conflict between pro- and anti-action workers.} Participants often highlighted workplace tensions between pro- and anti-action workers as a risk to their organizing effort. Rifts between workers led to conflicts and tensions in the workplace, made the organizing effort emotionally taxing, and---in extreme cases---involved co-workers reporting organizing activity to their supervisors \interview{2,3,7,8,16}. In some instances, we also observed emotion or tone as a threat to the collective effort, when potential union members observed a tone they perceived as "toxic" or otherwise off-putting \interview{3,14,27}. All of these also contributed to organizer burnout, putting the campaign at risk.
    \item \textit{Large-scale individual retaliation.} In many cases, employers took retaliatory action against a large number of organizers and supporters, thereby stunting or stalling the larger movement \interview{2,7,8,10,17,22,29}. We differentiate the types of individual retaliation below, but in these cases, the retaliation was widespread to the point that it had significantly crippling effects on the movement. 
\end{itemize}
\subsubsection{Risks to individuals}
\label{sec:indiv-risk}
Apart from risks to the collective effort, participants expressed several risks posed to themselves, their fellow organizers, and other action supporters as individuals. In nearly all of our interviews, participants themselves or their co-organizers were retaliated against by management in various ways that ranged from selective firing to poor performance reviews. This applied both to organizing campaigns that succeeded in their collective action, as well as those that did not.  These risks manifested at various stages of the campaign and affected different organizers differently, based on their identity and precarity of their jobs. 

\begin{itemize}
    \item \textit{Retaliation via job loss.} The most prevalent example of risks to individual organizers was the risk of retaliation as a result of their involvement in collective action. Almost every participant we interviewed indicated that some organizers were concerned about losing their jobs if their employers found out they were an organizer. One participant summarized: \bigquote{The risks here are retaliation, loss of job---those are the big ones ... Employers don’t have your best interests at heart, that’s not their job---you’re not expendable, you're just expendable to them.}{16}
    This risk was felt more intensely by organizers who had a stronger need for job security, including organizers from marginalized backgrounds, organizers not protected by the NLRA under U.S. law (i.e. contractors), or organizers relying on work visas to stay in the U.S. \interview{2,9,12,16,17,23,26}. Many campaigns explicitly experienced firings correlated with involvement in collective action \interview{3,7,8,11,14}. 
    \item \textit{Retaliation via project loss or damage to professional reputation.} Participants disclosed a range of mechanisms that employers used to retaliate against organizers, and not all them involved firing the workers. A few participants described organizers getting negative performance reviews after their involvement in collective action, despite consistent work ethic \interview{2,17,29}. Others described how their relationship with their manager changed significantly for the worse after they became a part of the organizing effort \interview{21,23}. Yet another described how a peer organizer suddenly saw their projects being cancelled, a clear mismatch with the positive feedback they had received for that project prior to the campaign going public \interview{14}.

    \item \textit{Loss of social reputation.}
    Some participants described a fear that their reputation among their peers would be damaged by their involvement with organizing efforts, and that their efforts would bring about ill will among co-workers \interview{2,11,14,17}. One participant described that \smallquote{it doesn't feel good to know that your co-workers are in a Slack you're not in, talking shit about you}{14}. One participant did not want to be \smallquote{written off as [an activist] person ... the crazy liberal... my co-workers `muting me'}{11}.
    
    \item \textit{The toll of emotional labor and burnout.} Nearly all of our participants expressed feeling burnt out, or expressed that their co-organizers burnt out, often from emotional labor \interview{2,3,12,14,22,23,27,29}. Participants described that collective action required intensive work in addition to their day jobs, and progress was often slow and frustrating. In addition, when the company actively employed union-busting tactics, organizing work became more difficult to manage \interview{3,22}. Even though a majority of participants were well aware from the beginning that collective action would be an uphill battle, many organizers said that they did not originally foresee just how exhausting the process would become.
\end{itemize}

\begin{table*}[!h]
  \centering 
  \begin{adjustbox}{width=\textwidth,center}
  \begin{tabular}{P{3.4cm}P{3.4cm}P{3.4cm}P{3.4cm}}
    \toprule
    \textbf{Needs and Goals} & \textbf{Risks} & \textbf{Decision spaces} & \textbf{Example decision}  \\
    \midrule
    Reach out to as many members as possible & Employer retaliation, employer finds out about general strategy & Use personal devices or company owned devices? (\Cref{sec:challenges}) & Initial contact on company owned devices, follow-up on personal devices \\
    \midrule
    Build trust and a sense of community within the organizing collective & Employer retaliation, employer finds out about general strategy & Focus on usable and familiar tools or tools with the best privacy features? (\Cref{sec:usability}) & Use tools that most co-workers are familiar with to ensure greater participation \\
    \midrule
    Free and open communication and collaboration between members of the collective & Lateral worker conflict, centralized power, employer surveillance & Access control: open or closed? (\Cref{sec:accesscontrols}) & Require participation and 1-1 conversations before granting access to a new member \\
    \midrule
    Creating a healthy organizing environment & Conversations stifled due to moderation; no community management leads to discomfort & How much moderation is needed for communication within the organizing unit? (\Cref{sec:communitymanagement}) & Creation of explicit norms about engagement within the organizing unit \\
    \midrule
    Communication of sensitive topics & Privacy risks, employer surveillance & Paper trail or no paper trail in online communication and collaboration? (\Cref{sec:papertrails}) & Using emphemeral messaging and video calls instead of text communication to avoid paper trails \\
    \midrule
    Increasing power of the collective within company & Employer retaliation & Should the collective go public within the company with their collective action? (\Cref{sec:safetyinnumbers}) & Go public within the company only after a certain percentage of employees pledge to support the action \\
    \midrule
    Gaining leverage over management to get them to respond to demands & Privacy risks, employer retaliation and surveillance & Should the collective go public on social media with their collective action? (\Cref{sec:socialmedia}) & Don't go public on the internet if privacy risks to individuals seem high \\
    \midrule
    Increase membership and participation in the collective & New members not onboarded well, organizer burnout, employer hires union-busting firm & Grow membership in the collective quickly or more gradually? (\Cref{sec:paceofgrowing}) & Grow membership gradually to avoid burning out existing organizers \\
    \midrule
    Ensure openness, transparency and amity in the collective & Privacy risks, employer retaliation & Withhold strategic information from the whole collective to prevent leaking? (\Cref{sec:navigatingvisibility}) & Withhold key information, but share other information widely to ensure a culture of trust and openness \\
    \bottomrule
  \end{tabular}
  \end{adjustbox}
    \caption{Answering \textbf{RQ1}: Illustrative examples of how the needs and goals, risks, mitigations and decisions interact.}
    \label{tab:results}
\end{table*}

\subsection{Responses and challenges for mitigating risks (\textbf{RQ1-c})}
\label{sec:responses}
In response to the various risks outlined above, participants took several kinds of measures to protect themselves and the collective. This involves decisions on the part of the organizers---some based on technical criteria for ensuring their privacy and others based on social relations within their company. Our participants describe several decisions they needed to take during their organizing effort. We outline the salient decision spaces below and summarize responses to \textbf{RQ1} in \Cref{tab:results}.

\subsubsection{Challenges in avoiding company-managed resources}
\label{sec:challenges}
One of the most common pieces of advice given to organizers is to stay off company hardware and software: \smallquote{[the union] told us to never log into [the organizing] Slack on our work laptops. ... use personal devices, never use work laptop, don’t talk about it explicitly in the workplace, be careful when walking about it}{16}. Organizers, especially those with more precarious jobs, tended to follow this advice very closely. However, workers often encountered practical constraints while trying to follow these guidelines to meet their goals. For instance, while remote, using software platforms provided by the company is often the easiest or only way to initiate contact with co-workers with whom organizers otherwise had no relationship. Below, we outline various techniques participants adopted to mitigate increased risks of organizing using company resources and accommodate various practical needs and individual constraints.

\begin{itemize}
    \item \textit{Avoiding company-managed software.} It is almost never the case that organizers have access to personal contact information for all their co-workers. Many participants reported using company-managed software for initial outreach---often in the form of a vague invitation to chat about work---and moving to non-company channels afterwards \interview{4,8,7,9,11,14,21,25,26}. One participant related: \bigquote{We were careful at first, then we became less careful because we needed to reach people in some way, and cold DM via work chat is the only way to reach this person.}{26} Some reported feeling safe using company-managed video conference software for initial conversations once the organization went remote, and one-on-one meet-and-greet-type video calls became normalized \interview{2,14,16,21}. Although video call metadata could be accessed by administrators, participants suspected their company could not feasibly record and store all video chats. Several participants also related that they encouraged individuals to access shared documents using personal accounts rather than corporate ones \interview{7,9,16,27}.
    
    \item \textit{Avoiding company-issued devices.} In many cases, participants reported that they avoided using company devices for any organizing activity. However, this was not always possible. For example, some organizers did not have personal devices other than their phones, and one bought a personal computer to organize \interview{2,14,20}.
    Many organizers also related the practical difficulty and low perceived marginal security in, for instance, entirely preventing organizers from accessing personal email or other accounts on their work devices, and instead focused on ensuring they did not access organizing documents or communications through work-managed accounts \interview{7,9,27}. One participant related: \bigquote{You cannot totally isolate the organizing experience and work from your life. You can try to flag where that intersects, but no one has a totally separate computer with totally separate accounts for organizing.}{9}
    
    However, many organizers were not always certain which of the activities on their company or personal devices were being monitored. As one participant described: \bigquote{We definitely understood that our employer could see anything that we do on our company laptops and we knew that was bad because our union told us to organize off of work stuff. But we didn’t know where the boundary was exactly.}{27}
    A few participants strongly suspected or explicitly knew that there was surveillance software installed on their work devices, and sometimes even on personal devices used for work \interview{21,25}. These participants were well aware of the risks of organizing using work devices, and described various ways of mitigating them: not scheduling meetings on their work calendar, only doing organizing work outside of regular work hours, and setting boundaries with co-workers in order to ensure they were able to follow these precautions.
\end{itemize}

\subsubsection{Usability and accessibility of organizing tools}
\label{sec:usability}

When deciding on which tools to use for communication between organizers as well as for collaborating on organizing efforts, one of the key decisions that participants highlighted was embracing tools that were already familiar to most workers.

A few participants described preferring tools like Signal because they felt safer using it. For instance, one mentioned their fellow organizers "love the feeling of security", and another related, "if journalists are okay with [Signal], then I am [too]" \interview{1,20}.

However, others found it more difficult to use Signal if it was not already heavily adopted at their workplace. Complex onboarding processes incurred a cost onto new members that made it more difficult for them to get involved. One participant summarized, \smallquote{using tech that people are comfortable with means there isn't an extra thing to learn. Tech already exists to help people work together, that’s one less place people have to learn}{22}. Another related, \smallquote{I've been in too many organizations that trade security concerns for usability. My focus is much more towards making sure people can be part of the organization, over theoretical security concerns}{5}. Many participants described trying encrypted tooling like Signal, Cryptpad, and Keybase, but eventually switching back to more familiar alternatives to allow for greater participation \interview{1,4,7,9,16,24,26}. A few organizers relayed that the financial cost was another practical constraint that limited their selection of tooling \interview{1,14}.

\subsubsection{Access controls}
\label{sec:accesscontrols}

In order to keep digital communication and collaboration secure for the collective, almost every participant reflected that they used \textit{access controls} with their software. Access controls refer to a specification about who has permissions to access a document or communication stream, such as a message group. If an individual is not a part of the list of people allowed to access a certain document or message group, they will not be able to open the document or see the messages in the group~\cite{smetters_how_2009}. Decisions about access controls had important ramifications for the organizing effort. 

Access controls also expose an inherent tension between democratic access and managing the risk of information being leaked to management. Limiting access to documents reduces the risk of leaks, but can also limit the scope of democratic participation by all members in all decisions.

On one hand, many participants were extremely wary of allowing broad access that might enable documents containing core strategies and other sensitive data to be leaked to management via unsupportive co-workers \interview{4,9,14,17,21,24}. On the other hand, several participants and others cited transparency, openness, and worker agency as important values when building the collective \interview{8,11,14,17,21,23,24,28}. One worker related, \smallquote{the more transparent you are, the more vulnerable you are. But we pushed for transparency because people trust it more. And what we are doing is trying to foster a organization of trust}{17}.
Another participant expressed that having data available to more people would enable easier delegation of tasks:
\bigquote{There were some levels of privacy that were really important ... I wouldn’t let anyone who hadn’t [shown support] to access those [documents]. I was the only one doing managing access because no one else had time. ... I wanted that to be available to more people but didn’t want to risk it getting out. That way, people would have been more involved ... and they’re an actual stakeholder}{21}
 
Maintaining an access control policy required dedicated organizers who were responsible for maintaining a list of who was allowed to view each set of documents or messages. Many participants reported expending significant effort updating the access control list to make sure it reflected recent changes in the structure of the organizing unit \interview{9,11,14,21,26}.

\subsubsection{Community management and moderation}
\label{sec:communitymanagement}

The goal of many organizing groups was often to build cohesion and a sense of trust between workers, something that was tangibly missing from the remote environment. Many participants related the importance of mutual empathy and trust as a way to feel safe within the group \interview{9,11,16,17,20,21,22}. One expressed that \smallquote{it was more about a sense of trust with other people that contributed to a feeling of safety than the platform}{9}.

Several participants emphasized the need to craft communication policies or community moderation guidelines to protect this sense of trust. These guidelines varied between collectives. Some participants worked to ensure continued good relations among members of the collective by having moderators de-escalate heated conversations \interview{8,11,14,22,24}. Some collectives had moderation to ensure discussion spaces were understood to be somewhat public so that members did not accidentally mention sensitive information about individuals without their consent \interview{1,9,14,28}.

Participants also described how a lack of rules of engagement within organizing spaces could lead to burnout within the organizing unit, stressing the magnitude of emotional labor that goes into community management \interview{4,9,11,14}. These participants all described a balance between having a opportunity to vent in the organizing space and maintaining a culture of inclusion and positivity---for example, when an organizing group was used excessively to vent about issues with management, a participant described, on one hand, feeling uncomfortable remaining in a space that was overly negative, but on the other hand, feeling a stronger sense of belonging and trust within a group that was trusted with this information \interview{4}.

\subsubsection{Being cautious about digital ``paper trails''}
\label{sec:papertrails}

One impact of the shift to remote organizing is that, unlike in-person conversations---which were heavily utilized by organizers for sensitive and other one-on-one conversations---online text-based discussion leaves a digital trace.

Some participants wished they had been more careful about early, private collective conversations, which were visible to later members after adding them to a group workspace with full history \interview{4}. Some participants set up certain communication channels to delete messages by default, or simply restricted sensitive conversations to ephemeral video calls or in-person discussions when possible \interview{9,14,20,21,22}. One participant indicated their general lack of trust in digital tools, where the default is to leave "paper trails" that could later be accessed by others \interview{21}.

We note that this concern is similar to, but slightly different from, whether using a particular chat application leaves message contents encrypted on the service provider. In fact, while some preferred the use of encrypted tooling, many participants related that encryption was not necessarily as important to their tool selection criteria as features like familiarity to workers, ephemeral messaging, or highly customizable access controls \interview{5,9,14,21,24}.

On the flip side, many organizers emphasized the necessity of keeping records of events like retaliatory or harassing messages or explicit retaliation by management. One participant shared that, \smallquote{we have a giant folder full of screenshots of crazy shit managers or executives say… Written stuff gets you in trouble…. I have a decent case for retaliation if they fire me}{23}. Several participants noted that their employer also often avoided documenting policies via chat or email, preferring to specify sensitive policies or discuss controversial items over video chat \interview{20,24}.

\subsubsection{Safety in numbers}
\label{sec:safetyinnumbers}

Throughout our interviews, we noticed an inherent tension between safety through tight individual privacy and safety through large collective action. In a few cases, for instance, the secrecy of an effort led to a lack of trust within the larger company and made it difficult to delegate tasks to new members \interview{9,21}. On the other hand, if they revealed their core organizers before they had enough support, their employer could retaliate against them and break apart the collective effort, as demonstrated in \Cref{sec:risk}.

Due to the myriad risks shouldered by individuals when organizing or participating in collective action, some participants mentioned their communication channels allowed participants to use pseudonyms during the earlier phases of organizing~\interview{1,4,9,16}.

Several participants mentioned the importance of publicity and strength in numbers to both the chances of campaign success and to collective safety \interview{2,7,20,24,26,27}. One elaborated, \smallquote{if we move together, there is very low retaliation}{20}. Advice received by one participant was that \smallquote{[it] turns out when you’re organizing, public support is the only support that matters}{2}. Another elaborated a mutual commitment organizers made with each other:

\bigquote{If anything happens to one of us, we will do anything, up to the point of work stoppage, to show our strength behind that person. ... Most people in the organizing committee have psychological safety by virtue of them being connected to so many people. ... the risk is much lower because there’s so many people ... they can’t just reasonably mysteriously fire all of us without it being a very big deal.}{26}

Another organizer related that many individuals were willing to "go public" with their support after a certain percentage of support was reached, and that individuals at different risk levels showed different levels of willingness to go public with support for the collective \interview{7,11,24}. Participants outlined several other signals of support used by members of the collective. For example, several participants reported using digital visual markers---such as emojis, custom zoom backgrounds and Slack profile pictures---to signal their solidarity with the collective \interview{8,11,17,22}. 

\subsubsection{Relationship to journalists and social media}
\label{sec:socialmedia}

In some cases, participants also reported taking public action on social media or leveraging news media to put pressure on the company to respond to their asks~\interview{1,8,11,23,24,25}. One organizer mentioned putting together internal media trainings to build a media presence without accidentally disclosing too much information about their efforts~\interview{1}.
We also note that companies did surveil individuals' public behavior, including having HR or managers hunting for anonymous participants in media interviews, or chastising organizers for their quotes in news articles or posts on social media~\interview{4,7,16,24,26}. A few participants mentioned that, even if individuals felt comfortable being public about supporting the collective at their workplace, many were uncomfortable being public on the Internet due to personal privacy concerns, as well as the risk of opening individuals to the threat of doxxing or other harrassment~\interview{16,23,26}.

\subsubsection{The pace of growing the collective} 
\label{sec:paceofgrowing}
Participants highlighted the tension between growing the collective quickly and the risks inherent in adding too many new members in a short period of time. One expressed wanting to move faster to counter a particular policy: \smallquote{I wanted to push it along really fast because I knew people were getting pushed back into the office [even though they felt it wasn't safe to return]}{21}. Many other participants expressed wanting to move quickly after going public, in order to prevent management from realizing further union-busting tactics \interview{2,7,17,22}. However, they felt that the timeline ended up being too rushed, as they received too large an influx of new members to properly on-board onto their community practices, often leading to organizer burnout.
A campaign that went on for too long also ended up demotivating organizers, cutting momentum, and burning out organizers: \smallquote{it’s hard to keep people engaged for months and months at a time}{26}. Organizers not only have to carefully balance the speed at which they grow the collective with their capacity for on-boarding new members, but must also watch out for long-term organizer burnout and management implementing union-busting strategies.

\subsection{Effect of the shift to remote work on organizing efforts (\textbf{RQ2})}
\label{sec:remote-work}
Earlier in this paper, we discussed the risks and mitigation strategies that labor organizers in tech adopt while organizing for collective action. Many of these were affected by the shift to remote work. Here, we synthesize some of the main impacts that going remote had on organizing efforts.

\subsubsection{Accessibility of organizing.}\label{sec:accessibility}

Moving from in-person to remote work made it easier to get involved in organizing for collective action in some cases, because it no longer required organizers to meet at a separate physical location. Instead, organizers could reach out to co-workers online. For example, a participant described that when a majority of organizing activity took place in person, it was difficult for remote organizers to get involved. This quickly changed after the company went remote \interview{18}. Another participant expressed that the remote environment made it easier for a neurodivergent organizer to engage with people \interview{28}. If the workforce is spread across  several different offices, remote organizing may even be a necessity. Participants expressed that remote work made it easier for them to engage with co-workers who were in different geographic locations \interview{6,11,18,19}. 

\subsubsection{The built environment of the remote workplace.} \label{sec:builtenv}

The changed environment of the workplace made certain physical security concerns obsolete, but raised new ones. Digital communication technologies made certain physical surveillance concerns, such as shoulder-surfing or eavesdropping, obsolete. It became easier to be sure that conversations about collective action could not be overheard, like they might be in a shared physical office space \interview{17,21,26}. A participant indicated that their organizing trainings had all been centered around how to have safe in-person conversations in a physical workplace \interview{13}.

As mentioned briefly in \cref{sec:challenges}, in their transition to remote, many participants' companies also helped facilitate conversations between workers on different teams via meet-and-greet-type events~\interview{4,5,17,24,29}. These same participants expressed that the normalization of cold-contacting co-workers helped them start forming a relationship with workers across the company, and made them feel slightly safer contacting co-workers about organizing efforts as well. In addition, the remote environment made it seamless to switch contexts depending on the sensitivity of the conversation: a conversation could be immediately moved from company Slack to a private Signal chat.

In many cases, the transition to remote work was relatively smooth, since many employers in the tech sector already had existing infrastructure to accommodate remote work before the pandemic. However, the ease of the transition carried important drawbacks. Corporate tracking software was already common before the shift to remote work, but the shift to fully remote or hybrid workplaces might magnify the incentives to increase or heighten surveillance. Some participants suspected or knew that their employer installed invasive spyware after the pandemic began \interview{1,4}.
\begin{table*}[!t]
  \centering 
  \begin{adjustbox}{width=\textwidth,center}
  \begin{tabular}{P{4.5cm}P{9.5cm}}
    \toprule
    \textbf{Factor of remote work} & \textbf{Impact on collective action}  \\
    \midrule
    Accessibility (\Cref{sec:accessibility}) & 
    \begin{minipage}[t]{\linewidth}
    \begin{itemize}[leftmargin=*,nosep,after=\strut]
        \item[\texttt{+}] Workers who were always remote can get more involved in collective action, compared to when the majority of their co-workers worked in-person
        \item[\texttt{+}] Organizers can more easily talk to their co-workers in different geographic locations 
    \end{itemize}
    \end{minipage}
    \\ 
    \midrule
    Built environment of the remote workplace (\Cref{sec:builtenv}) & 
    \begin{minipage}[t]{\linewidth}
    \begin{itemize}[leftmargin=*,nosep,after=\strut]
        \item[\texttt{+}] Conversations about collective action cannot be overheard, as they might be in a shared physical office space 
        \item[\texttt{+}] Easier to switch contexts depending on the sensitivity of the conversation
        \item[\texttt{-}] Corporate surveillance is made easier, as work done remotely is vulnerable to employer-managed spyware
    \end{itemize}
    \end{minipage}
    \\ 
    \midrule
    Forming and maintaining relationships with co-workers (\Cref{sec:relationships}) & 
    \begin{minipage}[t]{\linewidth}
    \begin{itemize}[leftmargin=*,nosep,after=\strut]
        \item[\texttt{-}] Building trust and strong social ties with co-workers online is difficult
        \item[\texttt{-}] Emotionally charged conversations become more mechanical
        \item[\texttt{-}] Lack of non-verbal cues makes it hard to manage relationships
        \item[\texttt{-}] Breakdown of work-life barriers means people receive work-related messages during all hours of the day
        \item[\texttt{-}] Some channels of communication cannot be muted due to company policy, which leads to the channel becoming a \textit{captive audience meeting} which the workers cannot quit
    \end{itemize}
    \end{minipage}
    \\ 
    \bottomrule
  \end{tabular}
  \end{adjustbox}
    \caption{Answering \textbf{RQ2}: Factors influenced by remote work and their positive (\texttt{+}) and negative (\texttt{-}) impacts on collective action.}
    \label{tab:resultsRQ2}
\end{table*}

\subsubsection{Impacts of going remote on the relational aspect of organizing.} \label{sec:relationships}

While going remote made certain types of outreach and communication more convenient and accessible, many participants agreed that it had major drawbacks as well. In nearly every interview, we heard that building trust and strong social ties with co-workers was more difficult online. Workers often mentioned relying on strong relationships they had built \textit{before} going remote as an important starting point for deciding whom to reach out to first about joining the collective \interview{12,18,21,24}. In general, participants expressed that the difficulty of building community and trust online meant they had to work much harder for that same level of trust~\interview{4,12,16,17,18,27,29}. One participant related an experience of another organizer on the extra labor of maintaining personable relationships online:
\bigquote{It’s harder to do this over a video call than it is in person, because in person you’re gonna see them in the office again in the real world, and it continues to humanize and endear you in person and you’ll continue to get to know each other more after the fact. When it’s a video call, it’s more structured and a little less humanized. You’re not gonna see them again unless you do more follow-ups and you engage with them.}{4}

Beyond its effects on individual relationship-building, going remote also had profound impacts on internal discourse related to collective action as a whole. Many participants indicated that emotionally charged or difficult conversations become more mechanical, and that the lack of non-verbal cues made it hard for many to manage on both text-based discussion channels and large video calls \interview{17,21,25,28,29}.

Some participants also noted that the increased reliance on workplace discussion fora, and subsequent breakdown between work and life barriers, meant that people were always receiving notifications about activity in work channels \interview{17,28}. Since remote workers operate on a variety of schedules, these emotionally charged asynchronous discussions bleed into all hours of the work day, further burning out employees and organizers alike. These participants related this experience to being in a constant \textit{captive audience meeting}, as they could not mute notifications from some of these channels.

\subsection{Other trends in our results}
\label{sec:other-trends}
In this section, we describe several other trends that we noticed that do not directly address our research questions, but were particularly striking and warrant future work.

\subsubsection{The role of identity in collective action.}
In our results, we describe our observations that identity and context affected the risk perceived by that individual; for instance, in \Cref{sec:indiv-risk} we describe that organizers from marginalized backgrounds, or organizers relying on work visas to stay in the U.S., felt various risks of retaliation more intensely. In some cases, it also became a way to express solidarity and ensure collective safety. In many interviews, participants expressed that organizers with marginalized identities often refrained from publicly associating with collective action when it presented a risk to them---and were often actively supported by their fellow organizers in doing so. Instead, they relied on other fellow organizers to carry out public-facing work. In these cases, organizers who felt less at risk of retaliation leveraged their privilege to ensure the success of the organizing effort \interview{8,12,21,29}.

Identity in some instances introduced a contentious dimension to organizing efforts, reflecting the troubled history of unions when grappling with issues of race and gender \cite{crain_labors_2001,hill_problem_1996}. In a few cases, participants mentioned employees vocalizing that the lack of Black workers at their company was one of their reasons for being supportive of the organizing effort \interview{2,12}. In both cases, co-workers interpreted this reasoning as an accusation that anyone who was not in favor of the collective action was also racist.

\subsubsection{The differing nature of contract and platform-worker risks and challenges.} \label{sec:platform-workers}
Although we noticed similar trends in our interviews with platform-workers, certain risks and challenges manifested in different ways due to the mediation of employment by a digital gig work platform, the ephemerality of platform labor, and the gaps in U.S. labor law when it comes to workers who are currently (mis)classified as ``contractors.''

At least one platform-worker organizer perceived a correlation between their organizing activity and their \textit{deactivation} from the gig work platform (in other words, getting fired): \smallquote{Every single member of our core organizing committee was deactivated ... The intention was to cause chaos to their operations}{19}. While the National Labor Relations Act (NLRA) at least in theory protects employees against retaliation for participation in union activity, the NLRA does not cover independent contractors, who are considered ``self-employed'' in the U.S.

While platform-worker actions share some similar challenges to the other collectives we heard about, there were a couple in particular that stood out as unique. For instance, all three participants related that institutional memory is a very large concern, as worker churn is very high. In addition, all of our participants found the collective not through an employer-managed communication application like Slack, but through social media and other Internet forums shared with other platform-workers. From the outset, the platform-workers we talked to tended to be open about the fact that they were organizing and did not try to hide their intentions from the companies. They speak to the press about their efforts and recruit members through social media advertisements. There is also a feeling that they are starting from scratch, due to the clear gaps in U.S. labor law when it comes to offering protections for contracted workers, or allowing companies to misclassify full-time workers as contracted or part-time workers. 

The evolving landscape of platform labor warrants further study, as purveyors of this algorithmically mediated labor adapt employment and retaliation techniques to escape the scrutiny of workers, journalists, and labor law. We see possible connections between historical analysis of mid-century union tactics, like data transparency, wage contestation, and strategic participation, to the issues faced by contemporary platform-workers \cite{khovanskaya2019tools}.


\section{Discussion}

Our results show that the specific nature of risks to privacy result in different technical and social approaches to risk management by labor organizers in tech. Threats from management tended to be addressed with traditional digital security practices, while threats from conflict among workers required sustained community management, strategies to build social trust, and strong social norms. In the latter case, organizers often had to work \textit{around} technology that did not fully meet their needs.

A striking theme in our interviews was the relationship between individual and collective privacy. The dominant modes of thinking about privacy operationalize the concept at the individual level, but this is insufficient for understanding how safety through individual and collective privacy are deeply intertwined. Our study suggests a need to broaden the notion of safety in privacy research to address the needs of collectives.

We also provide a set of design recommendations for digital communication and collaboration platforms. Although we caution that policy is not our area of expertise, we also provide a few suggestions for policymakers. These recommendations should not be seen as any kind of silver bullet---rather, they are suggestions for how the baseline needs of organizers in the tech sector could be better met. Finally, we conclude with areas where we identified a need for future research.

\subsection{Navigating visibility: Openness and transparency vs. collective security.}
\label{sec:navigatingvisibility}
Across our results, we noticed that one core tension organizers wrangle with is the desire to build an open, transparent, and trusting organization, while still grappling with both individual safety and collective data security. In particular, protecting the security of data from management requires the strategic withholding of information. To protect individual privacy, a collective might choose to allow individuals to use pseudonyms. To protect the privacy of the collective, they might operate in secret for as long as possible before going public, in order to defend against union-busting, and be selective about who should be granted access to certain documents or communication channels.

Participants noted that these withholding strategies can conflict with values of openness, transparency, and amity. When a collective intentionally revealed the identities of core organizers in an effort to appear more open, those organizers became targets of retaliation. On the other hand, when the collective withheld information about key decisions or organizers, it was difficult to build a culture of trust. Similarly, inviting non-supportive workers to collective action-oriented communication channels risks not only leaking campaign information to management, but also burns out organizers who must take on the role of community moderators and mediators. Organizers might do so anyways in order to prioritize transparency, and instead ensure (via moderation or other policies) participants in certain conversation channels understand that they are somewhat ``public'' spaces.

\subsection{Interaction between threat actors and risk management strategies}

When faced with risks to privacy during the organizing effort, we found that who the threat actor was had an effect on what responses the organizers took. Risks from the employer or management led to changes in digital practices, such as being cautious about digital paper trails and using personal devices. In these cases, it was assumed that as long as management did not have access to information about the organizing activity, the collective could remain safe. 

On the other hand, risks from co-workers led to mitigations of a different kind: in addition to digital practices such as adding access controls to organizing documents, these included more social and community-based approaches to risk mitigation, including dedicating significant organizing capacity to community management and gating access to resources on participation in conversations. We heard that the tools considered most secure (e.g., Signal) are not necessarily usable for large organized collectives. We confirm that usability was a key factor that drove technical decision-making: organizers gravitated towards tech that was familiar, convenient, and functional. Digital tools that are secure on paper but impede organizing can thus be counterproductive for individual and collective privacy. 

Our results show that organizers have a variety of ways of keeping themselves and their collectives safe against different risks, and these responses change based on who the threat actor is. In particular, we stress the centrality of community management and care work to the privacy of organizers involved in collective action. Many of the most salient privacy risks---and the ones organizers felt least prepared for---ultimately stemmed from the human aspects of collective action, such as lateral worker conflict and the emotional toll of employer retaliation. 

\subsection{Relationship between individual and collective privacy}
The relationship between individual and collective privacy that emerged from our interviews was particularly noteworthy. As expected, we did observe tension between being public with the organizing effort in order to build worker power and being discreet about the organizing effort in order to reduce risks of retaliation. Participants describe various ways in which they deal with these risks: by relying on personal connections when reaching out to new members, by relying on internal message boards and forums to decide which co-workers might be safer to reach out to, and by having organizers in less precarious positions taking public actions on behalf of the collective.

However, we found that there is also a sense in which individual and collective privacy are \textit{mutually constitutive}: when individuals feel a sense of psychological safety in the group, they are also often motivated to take on more organizing responsibilities. Likewise, the existence of the collective allows individuals to feel a sense of shared agency and provides a safe space for venting, criticism, and getting to know their co-workers through informal conversation.
Though developing a theoretical model for the relationship between individual and collective privacy is outside the scope of this current work, we see it as a promising direction for further research (Section \ref{sec:future_work}), one that is aligned with prior research that frames privacy and security as collective social and cultural practices \cite{dourish_collective_2006}.

\subsection{Limitations of modern privacy frameworks and comparison to other works}
It may be inadequate to retrofit existing models of information security onto the myriad privacy needs of labor organizers. In the past decade, computer security and security design communities have generally accepted the importance of  \textit{usable security} principles in order to counter security maximalism by grounding traditional information security principles in the real behavior and needs of users~\cite{sasse_usable_2005}. 

This work contributes to a vast body of research examining the digital security, privacy, and safety needs of specific at-risk populations~\cite{mcdonald_its_2021,havron_clinical_2019,owens_x201cyou_2021,simko_computer_2018,mcgregor_investigating_2015}. Like in many of these other studies, for labor organizers, digital security and safety decisions must support their primary goal (in this case, conducting a collective action successfully to influence policies at the workplace). The focus of threat modeling on protecting “information assets” is not adequately holistic for systems like this where information privacy is a need, but information is a byproduct rather than the primary source of value, such as grassroots activist collectives~\cite{aouragh_lets_2015}.  
For instance, in our work we find that organizers prioritized creating and maintaining relationships, protecting each other from burnout, and achieving particular organizational goals. The privacy of the information byproduct is still important (i.e. to be managed as digital paper trails), but may be secondary to these other needs.

Despite many shared risks and shared goals of labor organizers, due to the varying cultures and sizes of workplaces, non-contextualized security advice can be insufficient or oversimplifying. Our work demonstrates several contradictions and tensions navigated by organizers that is not easily addressed by straightforward security advice. For instance, in our work, we highlight the practical difficulty of following the most common digital security advice given to labor organizers to organize completely off company devices. We also highlight the inherent tension in building a transparent and inclusive organization while securing collective information from management; in our results, organizations that preferred transparency to security or vice-versa each faced their own unique challenges.
As another example, many organizations that were advised to use particular tooling (e.g. Signal, Keybase, Discord) often ended up migrating back to more commonly used platforms. Usability and accessibility were the primary factors in determining the long-term stability of tools and platforms by organizers, but we also note that tool or platform choice were secondary to community management decisions, in terms of their practical effects on realized collective or individual risks.

\subsection{Design recommendations}

We emphasize that, in accordance with our findings around the importance of community management, technology or design interventions cannot solve the many tensions and challenges that workers face while taking collective action. As prior work establishes, emergent digital platforms and technology design trends come from a long history of scientific management tooling that prioritize the control of labor power~\cite{khovanskaya2019tools}, and interventions in these technologies alone will not fundamentally alter workers' relationship to labor.

We prioritize suggestions to existing software that is familiar to, and commonly used by, labor organizers to communicate and organize information. Our results demonstrated that despite the fact that other platforms or tooling may have more desirable features, collectives will prefer to use tooling that they are familiar with to cut the labor cost of additional on-boarding. Below, we enumerate several privacy and usability design recommendations for digital collaboration tools to improve the experience of organizers. 

\subsubsection{Clear documentation on extent of admin access to organizational data}
Many organizers were confused about or did not know the extent to which administrators could surveil their activity on common organizational workplace software, such as Office 365, Google Workspace, Slack, and Zoom. Vendors of organizational or enterprise software should have public and accessible documentation on exactly which kind of data can be accessed by each user depending on their roles and permissions. Organizers often relied on members with IT expertise for this information to minimize the risk of reaching out to co-workers on company-managed software.

\subsubsection{Improved functionality on the free tier} Many of the communication and collaboration technologies preferred by organizers have limited functionality in their free tier. For example, the free tier of Slack deletes all messages prior to the most recent 10,000, making it difficult for collectives to access historical knowledge. Similarly, Slack and some other tools do not provide user group-based permissions in their free tier, a feature that makes managing access controls to communication channels much easier.

\subsubsection{Easily searchable archives} An integral part of maintaining institutional memory for collectives is having an archive of key decisions. With the shift to online tools for collaboration and communication, these decisions are increasingly taken on digital platforms. However, while these platforms enable dynamic conversations between users, they often do not focus as much on allowing users to create a cohesive archive of their communications, which leads to lapses in institutional memory.

\subsubsection{Flexible administrative options} Most online platforms allow individual users to become administrators for a group or communication channel, but often do not allow any collective decision-making process within the group itself. This is antithetical to the collective nature of labor organizing, which often aims to take decisions democratically. Adding collective decision-making capabilities would allow these digital tools to reflect the values of the collectives using them \cite{zhang_policykit_2020}.

\subsubsection{Better multi-account support}
Some organizers highlighted the difficulty of implementing the most common policy of sticking strictly to non-company-managed devices because they do not own multiple devices. Android's "Work Profile" isolation of work and personal applications and data is one option; however, usage of this feature requires that your employer uses Google Workplace, and that you own an Android device. Device and operating system vendors should prioritize and democratize access to these sandboxing features.

\subsubsection{Customizable and clear data retention policies}
In the transition from in-person organizing to online, organizers often desired to avoid digital paper trails of particularly sensitive conversations. However, many messaging and communication platforms do not allow this option. Customizable message retention policies should be a higher priority for vendors of large communication applications like Discord and Slack. The option of disappearing messages would allow collectives to talk more freely and reduce the burden on admins to make sure that sensitive messages are deleted after a given amount of time.

\subsection{Role of policy in collective action}
Although the focus of this research was on the relationship between digital technologies and the privacy practices of organizers, participants repeatedly expressed frustrations with the inadequacy of U.S. labor policy. We include some of these findings as potentially relevant for policymakers.

\subsubsection{Strengthening worker protections in the U.S} 
As we noted, some form of organizer retaliation was surfaced in nearly all our interviews. Many organizers expressed that they were not adequately protected by existing NLRA protections. The trend towards retaliation against organizers demonstrates a clear need for broader worker protections for organizers in the U.S. 

\subsubsection{Limits and disclosure requirements for employee surveillance}
There should be limits to which workers can be surveilled at work, especially as a growing number of workplaces are going remote. If a workplace does perform digital or physical surveillance, it should be required to fully disclose those forms of surveillance via a notice or other privacy policy.

\subsection{Future work}
\label{sec:future_work}

The results of our research suggest several areas for future work. Some of these concern populations that our inclusion criteria did not cover, but that we believe may share similarities with our results. In an age where more aspects of our lives are becoming mediated by digital platforms, we feel that recent changes in organizing practices deserve further study. Other directions for future work relate to particular phenomena related by our participants, such as organizing a platform-based workplace, or increasing and more pervasive surveillance by employers.

\subsubsection{Remote organizers outside the computing technology industry.}
Our participant inclusion criteria was focused on the computing technology industry due to researcher proximity to this industry, the recent surge of organizing in this particular industry, and the high incidence of remote-friendly policies since COVID-19. We suspect that many organizers driving the recent resurgence of labor movements in the U.S. \cite{kullgren_us_2021} will share similar experiences and challenges to the participants we interviewed. A larger study of recent remote organizing practices both outside and inside the computing technology industry would be extremely valuable.

\subsubsection{International workers.}
The scope of our research was limited to organizers in the U.S., yet many of the essential forms of labor that fuel the technology industry are primarily located outside of the U.S. \cite{raval_interrupting_2021, qadri_platform_2021}. In a world where such labor is unevenly distributed by geography, research that explicitly focuses on workers outside of the U.S., particularly in the Global South, is another important and promising direction for future work.

\subsubsection{Platform-workers.} 
Three of our participants were platform-workers. As described in \Cref{sec:platform-workers}, we found that their experiences organizing transient workers had unique risks and challenges compared to both full-time employees and contract employees with a designated workplace and regular job schedule. As the gig economy rapidly expands, we think this is another immensely important direction for research and study.

\subsubsection{Individual and collective privacy.}
We would like to see future research on privacy not only focus on the needs and social contexts of individual users, but also attend to the needs and contexts of the collectives of which they are members. Further, understanding how individual needs intersect with collective needs might provide a richer look into how people assess risk in practice. For example, would a person who identifies strongly with the collective be more willing to share their pay information, because the collective need for equity and transparency outweighs the individual risk of being identified? How does one's self-concept as an individual or as a member of a group ultimately shape how they conceive of safety in the first place, thus affecting their choices as they pertain to privacy?

\subsubsection{The growing use and purview of bossware and other employee monitoring software.}
A few of the participants we interviewed indicated concern with tracking software installed on their machines after their offices went remote. These experiences map to the reported spread and growing pervasiveness of employer surveillance since the COVID-19 pandemic pushed many offices to adopt remote-friendly policies \cite{ball_electronic_2021}. Our work focused on the perceptions of this software and subsequent reactions by rank-and-file organizers. Work from the other perspective would also be extremely useful---for instance, the study of how employee surveillance is changing in tandem is another promising direction for future work.

\section{Conclusion}
An energetic resurgence of the labor movement in the U.S. is currently underway. Workers across industries, workplaces, and occupations are fighting for more control over their working conditions and the applications of their labor, and tech workers are no exception. While digital technologies have provided the infrastructure for much of the communication within these movements, particularly during the pandemic, they also function as sites of tension between privacy and openness, security and risk-taking, the individual and the collective. Through a series of qualitative interviews with tech workers in the U.S. involved in collective action at their workplaces, we identify key themes in the challenges to organizers' privacy. 

First, organizers respond to a diverse set of risks and threat actors by adopting digital security practices as well as social, community-based mechanisms for protecting worker privacy. These latter techniques were particularly important in dealing with lateral worker conflict, and led to the prioritization of usability and familiarity above purely technical features when choosing platforms to support internal collaboration within the collective. Second, participants framed privacy as a construct that operated at both the individual and collective level: although there were tensions between these types of privacy, privacy at one level was not possible without privacy at the other. We pair these theoretical observations with design recommendations that can create a safer environment for workers organizing in the computing technology industry.

While, as academics, we believe it is important to understand how labor organizers in the computing technology industry interact with digital technologies to secure the privacy of themselves and their co-workers, we also emphasize that technology is only one component of a collective’s larger strategy to win. Whether the collective effort succeeds or fails is underdetermined when viewed solely through the lens of technology: instead, organizing is an ongoing, active process that is influenced by a host of exogenous factors. During the interview process, we were repeatedly awestruck at the resilience, tenacity, and agency that organizers demonstrated in the face of illegal retaliation, heated conflict, and unpredictable obstacles. In these stories, we find hope not only in hard-won successes like the recognition of unions, but also the efforts that fell short of their goals. Resistance is always an exercise in which success is not guaranteed. That the role of technology is but one element in a larger story of human agency in the labor movement is, in our view, a cause for hope.

\begin{acks}
We thank Arvind Narayanan and Andrés Monroy-Hernández for their detailed feedback on a draft of this paper, as well as their feedback and support throughout the project. We also thank the three anonymous reviewers for CSCW 2022 who reviewed our paper and gave us actionable feedback, as well as the 29 labor organizers who generously shared their time and experiences during interviews. Sayash's work is supported by NSF grant CHS-1704444. Any opinions, findings, and conclusions or recommendations expressed in this material are those of the author(s) and do not necessarily reflect the views of the National Science Foundation.
\end{acks}

\bibliographystyle{ACM-Reference-Format}
\bibliography{references, references-1, references-addition}

\appendix
\addcontentsline{toc}{section}{Appendices}
\section*{Appendix}
\section{Results from the optional demographic survey}
\label{sec:survey_results}

  \begin{tabular}{ll}
    \toprule
    \textit{Race} & \textit{\#respondents}     \\
    \midrule
    White & 9 \\
    Black or African-American & 2 \\
    Asian or Asian-American	& 5 \\
    Native American/American Indian/Alaska Native & 1 \\
    Native Hawaiian/Other Pacific Islanders	& 0 \\
    Other & 0 \\
    \midrule
    \textit{Hispanic, Latino, or Spanish origin} & \textit{\#respondents} \\
    \midrule
    Yes & 0 \\
    No & 15 \\
    \midrule
    \textit{Gender} & \textit{\#respondents} \\
    \midrule
    Man & 4 \\
    Woman & 9 \\
    Non-binary/non-conforming & 2 \\
    \midrule 
    \textit{Age} & \textit{\#respondents} \\
    \midrule
    <25 & 4 \\
    25-34 & 8 \\
    35-44 & 3 \\
    45-54 & 0 \\
    55-64 & 0 \\
    65+ & 0 \\
    \midrule
    \textit{Highest level of education completed} & \textit{\#respondents} \\
    \midrule
    Less than high school	& 1 \\
    High school	& 1 \\
    Some college	& 2 \\
    College	& 11 \\
    \midrule
    \textit{Immigration status} & \textit{\#respondents} \\
    \midrule
    U.S. Citizen & 14 \\
    Green Card Holder & 0 \\
    On a visa & 1 \\
    \bottomrule
  \end{tabular}

\section{Demographic survey instrument}
\label{sec:survey_form}

\begin{enumerate}
    \item Which of the following describes your race? (click all that apply) 
    \begin{enumerate}
        \item White
        \item Black or African-American
        \item Asian or Asian-American
        \item Native American/American Indian/Alaska Native 
        \item Native Hawaiian/Other Pacific Islanders
        \item Some other race, specify:
        \item Prefer not to answer
    \end{enumerate}
    \item Are you of Hispanic, Latino, or Spanish origin, such as Mexican, Puerto Rican or Cuban?
    \begin{enumerate}
        \item Yes
        \item No
        \item Prefer not to disclose
    \end{enumerate}
    \item Gender (choose all that apply)
    \begin{enumerate}
        \item Man  
        \item Woman  
        \item Non-binary/non-conforming
        \item Prefer not to disclose
    \end{enumerate}
    \item Age
    \begin{enumerate}
        \item < 25 
        \item 25-34 
        \item 35-44 
        \item 45-54 
        \item 55-64 
        \item 65+
        \item Prefer not to disclose
    \end{enumerate}
    \item Highest education level completed 
    \begin{enumerate}
        \item Less than high school
        \item High school 
        \item Some college
        \item College
        \item Prefer not to disclose
    \end{enumerate}
    \item What is your immigration status
    \begin{enumerate}
        \item U.S. Citizen
        \item Green card holder
        \item On a visa
        \item Prefer not to disclose
    \end{enumerate}
\end{enumerate}

\section{Interview instrument}
\label{sec:interview_instrument}

{}

\begin{enumerate}
\tightlist
\item
  {Could you tell me a bit about where you work(ed) and what your role
  is?}
  \begin{enumerate}
\tightlist
\item
  {Were you a full time employee or contract worker?}
\end{enumerate}
\end{enumerate}

{}

\begin{enumerate}
\setcounter{enumi}{1}
\tightlist
\item
  {How did you get involved in organizing with your colleagues at your
  company? }
  \begin{enumerate}
\tightlist
\item
  {Looking for: Information about their motivations for organizing,
  their experience in being reached out to, what platforms were they
  reached out on}
\item
  {What was your motivation for getting involved in organizing efforts?}
\item
  {How did you get in touch with the collective?}
  \begin{enumerate}
\tightlist
\item
  {{[}If digital platform{]}}{~Which platform were you reached out on?}
\item
  {Did the choice of platform affect your perception of safety and
  privacy? How?}
\end{enumerate}
\item
  {What is your current level of involvement with the collective?}
\end{enumerate}
\end{enumerate}

{}

\begin{enumerate}
\setcounter{enumi}{2}
\tightlist
\item
  {Can you walk me through a specific organizing activity or a
  collective action campaign that you were recently involved in? }
  \begin{enumerate}
\tightlist
\item
  {{[}Examples: membership drive, reaching out to co-workers to ask them
  to join, starting a signature campaign, attending a group meeting,
  public campaigns to pressure employers.{]} }
\item
  {How many people were involved, and what was the outcome?}
\item
  {Looking for: A transition to the discussion about the communication
  practices, especially digital communication practices.}
\end{enumerate}
\end{enumerate}

{}

\begin{enumerate}
\setcounter{enumi}{3}
\tightlist
\item
  {How did you communicate with fellow organizers during this period? }
  \begin{enumerate}
\tightlist
\item
  {What digital tools or platforms did you use for communication with
  your fellow organizers and how?(For example iMessage, Signal or
  Whatsapp)}
  \begin{enumerate}
\tightlist
\item
  {{[}if yes{]}}{~Which digital platform did you use and why did you
  decide to use this platform? Who made the call for the platform use?
  What considerations did they have in mind?}
\item
  {{[}if no{]}}{~What alternatives did you use to communicate and why?}
\end{enumerate}
\setcounter{enumi}{1}
\tightlist
\item
  {What digital tools or platforms did you and your fellow organizers
  use for collaboration, for e.g. google docs? }
  \begin{enumerate}
\tightlist
\item
  {{[}if yes{]}}{~Which digital platform did you use and why did you
  decide to use this platform? Who made the call for the platform use?
  What considerations did they have in mind?}
\item
  {{[}if no{]}}{~What alternatives did you use to collaborate and why?}
\end{enumerate}
\item
  {Did you use other, non-digital/analog forms of communication?}
\end{enumerate}
\end{enumerate}

{}

\begin{enumerate}
\setcounter{enumi}{4}
\tightlist
\item
  {What risks do you perceive when you are organizing? }
  
\begin{enumerate}
\tightlist
\item
  {How do you manage these risks along with your fellow organizers?}
\item
  {How do you manage risks when communicating and collaborating over
  digital platforms }{{[}reference the platforms they mentioned in 4.a.i
  and 4.b.i{]}}{?}
\item
  {What privacy and security practices do you and your co-workers follow
  when organizing in your workplace? How did you arrive at these
  practices?}
\item
  {Can you walk me through a specific incident where you or one of your
  fellow organizers felt threatened or faced retaliation due to the
  involvement in organizing? What happened? How did you manage this
  situation?}
\item
  {What are the parts of digital communication on }{{[}platforms
  mentioned in 4.a.i and 4.b.i{]}}{~that make you feel safe? Unsafe?}
\item
  {What were your relationships with co-workers --- whom you were
  organizing with and otherwise?}
\end{enumerate}
\end{enumerate}

{}

\begin{enumerate}
\setcounter{enumi}{5}
\tightlist
\item
  {How do you increase membership in your collective? Are you involved
  in reaching out to other employees who are not yet a part of
  }{{[}collective action org{]}}{?}
  
\begin{enumerate}
\tightlist
\item
  {{[}If yes{]} }{How do you reach out to employees whom you are trying
  to organize?}
\item
  {{[}If an organizer since before the pandemic{]}}{~How did the
  pandemic affect the procedures for reaching out to employees?}
\item
  {{[}If digital platform{]} }{Which digital platform do you use to
  reach out? Who decides on the platform and protocols? What factors
  influence their decision?}
\item
  {How do you decide whom to contact and organize with in the workplace?
  What factors affect this decision (personal relationships, the
  employee's position in the workplace)?}
\item
  {What are the biggest risks when reaching out to co-workers (e.g.
  employees talking to management? Losing your job? Increased
  surveillance? Retaliation?)? Can you think of an example when any of
  these risks ended up affecting you or your fellow organizers?}
\end{enumerate}

\end{enumerate}

{}

\begin{enumerate}
\setcounter{enumi}{6}
\tightlist
\item
  {Who is aware of your efforts at organizing?}
  \begin{enumerate}
\tightlist
\item
  {Do your managers know about your efforts at organizing with your
  fellow workers? }
\item
  {{[}If yes{]} }{Was this intentional? How did they find out?}
\item
  {{[}If no{]} }{What steps did you take to ensure that management does
  not find out about organizing efforts?}
\item
  {Do you perceive your employer to be passively or actively monitoring
  your activity in the workplace? Did this change because of your status
  as an organizer?}
\end{enumerate}
\end{enumerate}

{}

\begin{enumerate}
\setcounter{enumi}{7}
\tightlist
\item
  {{[}If organizing since before the pandemic{]}}{~What part of your
  organizing practice has changed the most during the Covid-19
  pandemic?}
  \begin{enumerate}
\tightlist
\item
  {Can you walk me through a specific example of a change that took
  place in your digital communication and collaboration collaboration
  practice with your fellow co-workers?}
\end{enumerate}
\end{enumerate}

{}

\begin{enumerate}
\setcounter{enumi}{8}
\tightlist
\item
  {Are there other stories from other companies that have had an
  influence on the way you think about unions and organizations?}
  
\begin{enumerate}
\tightlist
\item
  {Who is in the bargaining unit for your specific team? Who was on the
  organizing committee {[}if there was some such structure in the
  organizing effort{]} and what were their backgrounds?}
\end{enumerate}

\end{enumerate}

{}

\begin{enumerate}
\setcounter{enumi}{9}
\tightlist
\item
  {Is there anything that you want us to know that we did not ask
  about?}
\end{enumerate}

{}

\begin{enumerate}
\setcounter{enumi}{10}
\tightlist
\item
  {Those are all the questions I have, thank you so much for your
  participation in this interview. }{{[}2nd interviewer{]}, }{do you
  have any questions to ask }{{[}interviewee{]}}{?}
\end{enumerate}

{}

\begin{enumerate}
\setcounter{enumi}{11}
\tightlist
\item
  {{[}snowball sampling{]} }{Who else should we interview?}
\end{enumerate}
\end{document}